\documentclass[11pt,twoside]{article} 
\usepackage{adassconf}

\newcommand{\pu}[2]{\ensuremath{#1\,\mbox{#2}}}

\paperID{O10.4}
\begin{document}

\nopagebreak
\title{ ECSS in the eXtreme}

\author{W.~O'Mullane, J.~Hoar, U.~Lammers} 
\affil{European Space Astronomy Centre, Avda. de los Castillos s/n
E-28692,  Villanueva de la Canada (Madrid) }

\contact{William O'Mullane}
\email{William.OMullane@sciops.esa.int}

\paindex{ O'Mullane, W.}
\aindex{ Lammers, U.}
\aindex{ Hoar, J.}


\begin{abstract}
The ESAC Gaia team engages in a form of eXtreme programming while 
the DPAC will follow a series of six month development cycles 
modeled on this approach. 
As a project within the European Space Agency 
 the European Committee for Space
Standardization (ECSS) standards are required. We present the bringing together of these realms.
\end{abstract}

\section{Introduction}
ECSS standards may be difficult to understand and without good tailoring are not well suited to 
the development of science processing software. Here we tell the DPAC story so far.

\section{The Gaia Satellite and Science} 
The Gaia payload consists of three distinct instruments for astrometric,
photometric and spectroscopic measurements, mounted on a single optical bench.
Unlike HST and SIM, which are pointing missions observing a preselected list of
objects, Gaia is a scanning satellite that will repeatedly survey in a
systematic way the whole sky. 
The main performances of Gaia expressed with just a few
numbers are just staggering and account for the vast scientific harvest awaited
from the mission: a complete survey to 20th magnitude of all point sources
amounting to more than one thousand million objects, with an astrometric
accuracy of 12--\pu{25}{$\mu$as} at 15th magnitude and \pu{7}{$\mu$as} for the
few million stars brighter than 13th magnitude; radial velocities down to 17th
magnitude, with an accuracy ranging from 1 to \pu{15}{km\,s$^{-1}$}; multi-epoch
spectrophotometry for every observed source sampling from the visible to the near IR.

Gaia is being constructed by EADS Astrium under contract from ESA and is scheduled for launch 
at the end of 2012. It is to spend five years at L2 carrying out its survey. The 35 meter antenna in
Cebrerros Spain (an occasionally New Norcia) will be used to downlink about 100 Terabytes of data.
 
\section{The Astrometric Global Iterative Solution (AGIS)} \label{sect:agis}
The astrometry of Gaia is calculated from the multiple observations in different directions 
of the five years of the mission. The approach to this data reduction is a block iterative 
robust least squares fitting of the data as presented at the last ADASS (O'Mullane et al. 2007). Briefly 
objects are matched from the successive scans, attitude (three dimensional orientation of the satellite) 
and calibrations are updated. Next the object  positions are solved followed by higher order terms usually referred 
to as global parameters (such as PPN $\gamma$). This reduction  provides us with :
observed (proper) directions to a subset of {\em well-behaved} (primary) sources, 
attitude of the instrument as function of time and 
transformation from field angles to pixel coordinates.
This requires selecting   $10^8$ of the $10^9$ sources as input to the process and treating, iteratively,
the $10^10$ transits.

\section {Is development of Science software different ?}\label{sect:scidiff}
\begin{figure}
\plottwo{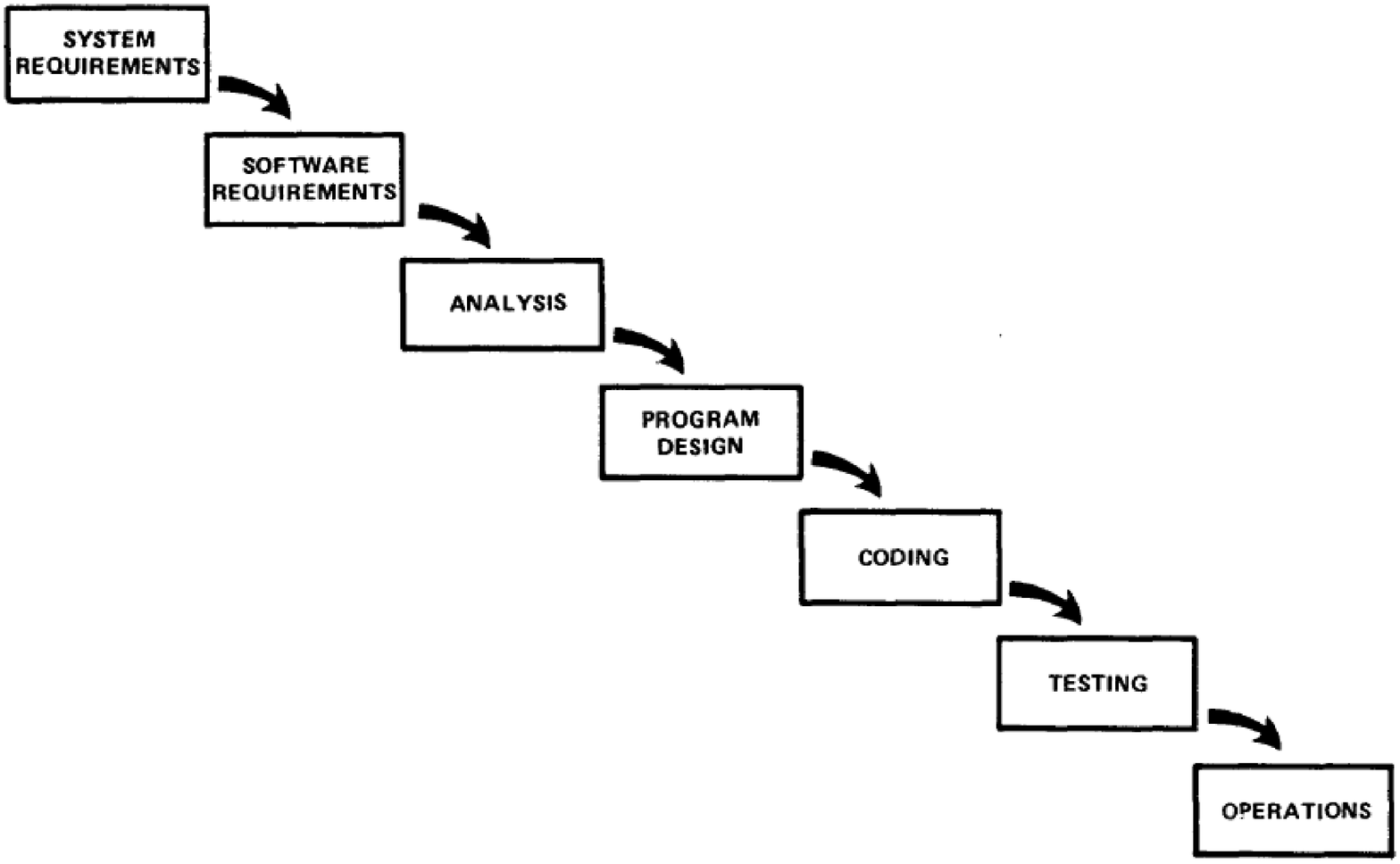}{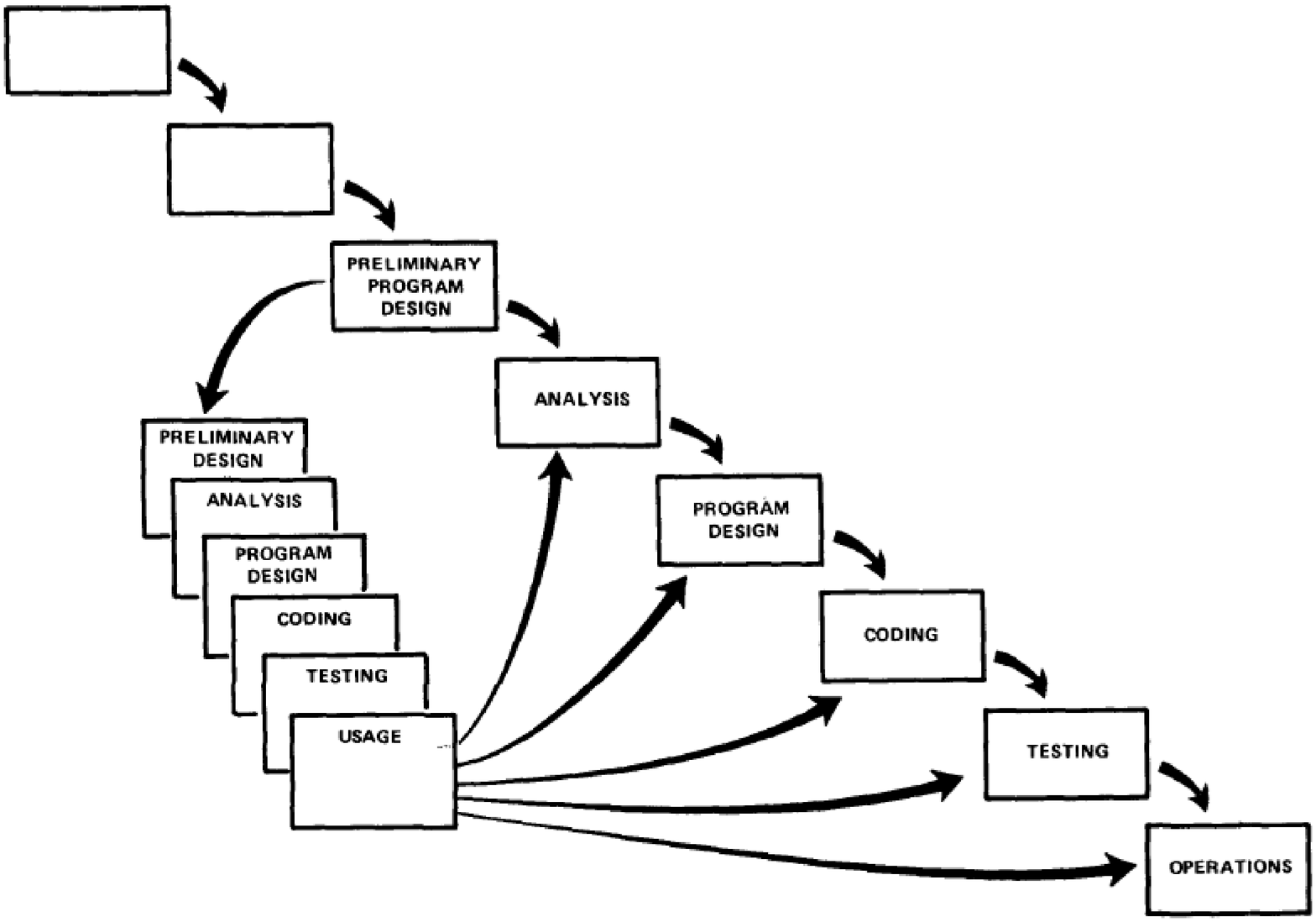}
\caption{Royce's waterfall model (left) reproduced from Royce 1970. This is held
to be the original waterfall approach but it is a scheme which Royce's paper
claims is flawed. The right figure appears later in the same paper.}
\label{fig:waterfall}
\end{figure}

The author conducted a survey over one year of many large science developments 
{O'Mullane 2005}. This proved to be  a most interesting study and concluded
that science software development is indeed different to traditional software 
development due to the funding structure and general approach to leadership.
This is in any case quite different to a satellite development project. Yet large 
organisations still wish to see developments like Gaia in a more traditional waterfall 
model. The approach to development within DPAC of using cycles
and always having some working software leans much more toward the {\em Agile}
techniques discussed here than traditional project management. Indeed these days
there are few if any software companies remaining who follow the waterfall
approach to development. The most lamentable thing of all is perhaps that
 Royce,attributed with the waterfall model,
showed the waterfall approach in Royce 1970  (see Left Fig.\ref{fig:waterfall}) as
an example of the flawed way to do 
software development. Later in the same paper Royce (see Right Fig. \ref{fig:waterfall})
he argues one should "do it twice". 

He points out the flaw in that it is a long time before
implementation starts and real problems are seen. These real problems may
fundamentally change requirements. All of these ideas
are in line with {\em Agile} approaches. Our intention is to do it far more than
twice - about ten times for DPAC. Even the preliminary AGIS 
went through several iterations before it was ready. 

The perhaps the fundamental difference between science software development and
satellite development is perhaps the risk element. Often for the software we may
not know exactly how a particular problem will be solved or indeed if it may be
solved. It would be a huge mistake to discard certain solutions simply to make a
good plan (which is after all what most managers would like to see). Let us take
some examples.

 \section{eXtreme development approach}
 No we do not all huddle around a single computer at ESAC but we have taken some useful ideas from the
 eXtreme programming field such as presented by Beck 1999. Planning is done on a monthly basis,
 stories are written and costed in points (where $ 1 point \approx 1/2 day$). The cost is agreed as a team 
 not set by {\em a manager}. Team members are also given points with which they {\em buy} the stories they wish to work on.
 Some stories of course must be {\em bought} first but generally team members should be able to have
 some story they would {\em like} to do in addition to the stories they {em must} do. All stories are entered in
 the XpTracker, a Twiki plugin designed for eXtreme programming.

 Code, in general, is owned by the group - any developer may {\em improve} code. Code reviews are carried out 
 frequently as a group where developers modify code and comments as well as  raise issues. In this way we try to keep
 code in adherence with he agreed coding standards as well as keeping documentation valid and up to date.
 Such an open system only works with excellent test harness in place. We use JUnit to write tests and 
 place emphasis on good test coverage (looking for 80 to 100\%). CruiseControl  is employed to regually check out code, build it and run
 all of the JUnit tests. Developers must keep tests passing when any modification is made to the code. The AGIS system ,for example, 
 contains 100K lines of Java, 30K lines of test code and 90K lines of comments. 

\section{European Cooperation for Space Standardization (ECSS)}
The European Space Agency (ESA) generally is more involved with satellite construction which follows
a waterfall model of development. Indeed for hardware production there is little choice. Springing from
this background the ECSS standards tell us how we should do a project, what documents should be
produced when and, to some degree, what they should contain. The standards come as a long list of books for example :
ECSS-E-10B	Requirements ,
ECSS-M-30A	Phasing,
ECSS-M-40B	Configuration Management, 
ECSS-Q-20B	Quality Assurance, etc.

The also come with a set of reviews which should be held which hark back tot he waterfall they are
the System Requirements Review, the Preliminary Design Review, the Critical Design Review, the Qualification Review
and the Acceptance Review. ECSS is considered applicable also to the ground segment software which includes the
science processing software.

\begin{figure}
\epsscale{0.8}
\plotone{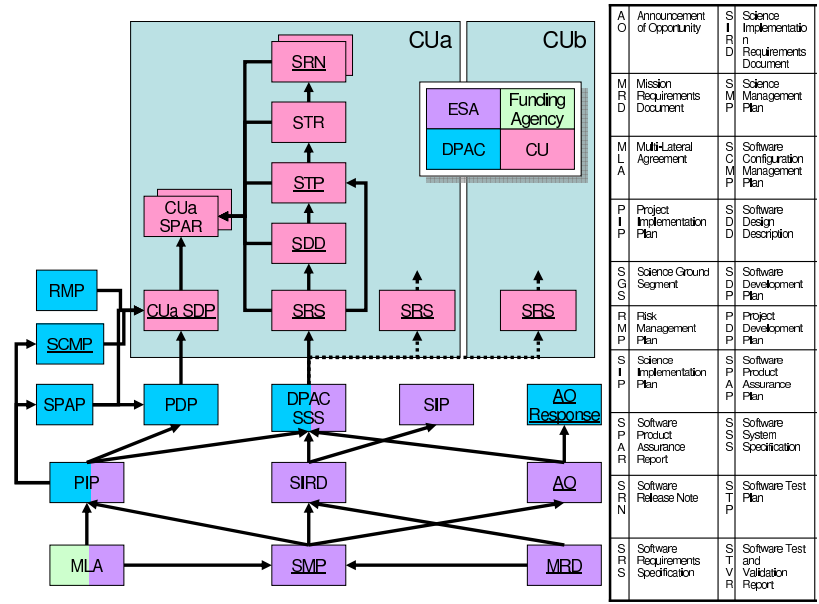}
\caption{Gaia DPAC document tree matching ECSS} \label{fig:doctree}
\end{figure}

In reality ECSS is very flexible - perhaps too flexible, it must be tailored to a particular project. Clearly some of
the ECSS is not applicable to software production so a {\em tailoring} must be undertaken. This outlines which parts of
ECSS will be used for the project and define to some extent the document tree (see Figure \ref{fig:doctree}). 

\section{Conclusion}
The usefulness of an eXtreme programming approach has been demonstrated for the AGIS software at ESAC. In general this
sort of approach is very well suited to scientific software development. If this will 
work in the scaled up manner for DPAC does remain to be seen. Every project needs documentation and with a little
effort the ECSS has been {em tailored} and a useful set of {\em standard} documentation outlined. Templates for 
these documents have been created. There remains considerable pressure toward more 
{em traditional} techniques for DPAC. Daily reading of Dilbert is recommended for sanity.


\begin{references}
\reference {{B}eck, Kent,Addison-Wesley 1999, Extreme Programming Explained: Embrace Change }
\reference {{O'M}ullane, William et al, Proceedings of ADASS XVI 2007, {G}aia {D}ata {P}rocessing {A}rchitecture},
\reference {{O'M}ullane, William,  Study Report 2005, A  study of {S}ome {L}arge {S}cientific {P}rocessing {S}ystems and {A}pplicabiliy to {G}aia},
\reference {{R}oyce, Winston, Proceedings of IEEE WESCON August 1970, {M}anaging the {D}evelopment of {L}arge {S}oftware {S}ystems},
\end{references}
\end{document}